\documentclass[aps,show pacs,twocolumn,amsmath]{revtex4-1}
\usepackage{amsmath}
\usepackage{hyperref}
\usepackage{graphicx}
\usepackage{float}
\usepackage{color}
\usepackage[normalem]{ulem}

\begin{document}
\title{Optimal first arrival times in L\'evy flights with resetting}
\author{\L{}ukasz Ku\'smierz}
\affiliation{Marian Smoluchowski Institute of Physics, 
Jagiellonian University, ul. \L{}ojasiewicza 11, 30-348  Krak\'ow, Poland}
\affiliation{AGH, Department of Automatics and Biomedical Engineering, 
Al. Mickiewicza 30, 30-059, Krak\'ow, Poland}
\author{Ewa Gudowska-Nowak}
\affiliation{Marian Smoluchowski Institute of Physics, 
Jagiellonian University, ul. \L{}ojasiewicza 11, 30-348  Krak\'ow, Poland}
\affiliation{Mark Kac Complex
Systems Research Center, Jagiellonian University, Krak\'ow, Poland}

\begin{abstract}
We consider the diffusive motion of a particle performing a random walk 
with L\'evy distributed jump lengths and subject to a resetting mechanism 
bringing the walker to an initial position at uniformly distributed times. 
In the limit of an infinite number of steps and for long times, the process 
converges to super-diffusive motion with replenishment. 
We derive a formula for the mean first arrival time (MFAT) to a predefined 
target position reached by a meandering particle and we analyze the efficiency of 
the proposed searching strategy by investigating criteria for an optimal
 (a shortest possible) MFAT.
\end{abstract}
\maketitle
\section{Introduction}

Limited random walks, with sudden termination of a trajectory are frequently 
analyzed in descriptions of motion in porous media, 
biological tissues, composite materials, and dynamic networks and of extreme, 
catastrophic events like gambler's ruin, chemical reactions, and species 
extinction~\cite{redner2001guide,Novikov,Sims,Harik1999,Zhu}. 
Quite often, however, the absorption events and disappearance of trajectories 
are followed by resets or restart activities of the system, e.g. the
relocation of searching paths in animal foraging, the seeking for target location by 
repair proteins or the returning to the initial position after an unsuccessful 
search of the address by an individual lost in 
a vast city \cite{benichou2011intermittent,mendez2013stochastic,majumdar2011resetting}.

A random walk with restart is also known as a graph mining technique widely 
used in the machine learning community for page-ranking or web search models
 and cryptology \cite{Galbraith,best-friends,chain-letters,tong2006fast}
In this approach the frequency of visits paid to a given node can be analyzed 
as a random walk on a graph. It is described as an ordered sequence of 
visits to vertices with a source (initial) vertex probability $\vec{p}_0$. 
For Markov chain models of transitions between subsequent locations on 
a graph described by a matrix $\Pi$, reset events inject additional randomness 
to the walk $\vec{p}_{i+1}=(1-c)\Pi\vec{p}_i+c\vec{s}_i$ with $c$ being 
the probability of resetting per step and $\vec{s}_i$ representing an arbitrary 
probability vector added at resetting. 

Many intriguing facets of the process in which a Brownian 
particle is stochastically reset to its initial position with a constant rate 
have been investigated by Evans and Majumdar \cite{majumdar2011resetting}. 
The stationary state of such a process has been shown to be described by 
a non-Gaussian distribution which, due to a non-vanishing steady state current 
directed towards the resetting position, violates the detailed-balance condition. 
The temporal relaxation towards this nonequilibrium steady state has been 
shown to exhibit a dynamical transition \cite{majumdar2015dynamical}.
Moreover, it has been proved \cite{majumdar2011resetting} 
that there exists an optimal resetting rate that minimizes the average hitting 
time to the target. Extensions to space depending rate, resetting to a random 
position with a given distribution and to a spatial distribution of the target 
have been also considered in Ref.\cite{Evans2011}. Brownian diffusion 
in external potentials have been further analyzed in a recent study by 
Pal \cite{Pal}. 

In a somewhat different context, similar random walks with stochastic resets 
have been analyzed by Durang et al. \cite{Durang} who posed the problem 
of interacting particles subject to a stochastic return to the initial configuration 
in the coagulation-diffusion process. The particles perform random hoppings to 
nearest-neighbour sites such that upon the encounter of two particles, 
the arriving particle disappears. The stochastic reset is described then by 
a given set of probabilities for having some consecutive empty sites.
A Markov monotonic continuous time random walk model in the presence of a 
drift and Poisson resetting events has been addressed in an elegant work by 
Montero and Villarroel \cite{Montero}, who derived general formulas for the 
survival probability and the mean exit time. 

While most of the works related to random walks with resets is based on 
continuous and discretized version of a Wiener process, 
relatively few studies have been devoted to resetting accompanying generalized Wiener 
motion with discountinuous L\'evy jumps.
L\'evy flights and L\'evy walks \cite{shlesinger1986levy} 
have been claimed to be observed in many foraging animal species 
\cite{viswanathan1996levy,viswanathan1999optimizing,atkinson2002scale,
ramos2004levy,brockmann2006scaling,reynolds2007free,reynolds2007displaced,
sims2008scaling,brockmann2010following,rhee2011levy}, which has 
led to theoretical analysis showing an optimality 
of L\'evy flights or L\'evy walks in different setups 
\cite{viswanathan1999optimizing,raposo2003dynamical,reynolds2005scale,
lomholt2005optimal,lomholt2008levy,zhao2015optimal}. 
The summary of those results 
can be found in a recently published book \cite{viswanathan2011physics}.
The optimization of a mean first passage time (MFPT) in 
a discrete time model of L\'evy flights with stochasting resetting 
has been addressed in Ref.\cite{kusmierz2014first}, 
where it has been shown that the optimal parameters admit jumps 
(i.e. discontinuous changes) as functions of a distance to the target. 
Hereafter, by analyzing statistics of first arrival times (FAT) of the 
continuous time version of the model, we demonstrate parameter-dependent transition between the optimal Gaussian 
and non-Gaussian search strategies.

In this work we concentrate on a variant of the model, 
in which a one-dimensional 
jump-like searching process with resetting events is analyzed as a renewal 
Markov model with L\'evy jumps.
We assume that a random searcher starts its motion at $x_0=0$ and tries to 
find the object located at some position $x$. The walker does not memorize its 
former locations, and the steps undertaken at any instant in time are 
statistically independent and drawn from a symmetric stable distribution with 
a stability index $\alpha\in(0,2]$. Furthermore, at random times following the 
Poisson point process, 
the searcher decides to instantaneously reset to the initial position. 
We derive an expression for the transition probability density of such process, 
analyze the existence and character of the long-time stationary distribution 
and discuss optimal conditions for the mean first arrival time (MFAT).

The paper is organized as follows: Section II introduces the model and 
discusses the structure and stationary solutions of evolution equations for 
corresponding probability distribution functions. The mean first arrival time 
in the model is introduced in Section III and its optimization is further 
analyzed in Section IV 
which presents the most important results of this work. 
We summarize the paper and add conclusions in Section V.

\section{Time evolution and transition probability} 

We start with an analysis of the integral equation that governs the 
evolution of the probability density function 
for the process $\{X(t),t\geq 0\}$:
\begin{eqnarray}
W(x,t|x_0,t_0)dx\equiv\nonumber \\
 \text{Prob}\{x<X(t)\leq x+dx|X(t_0)=x_0\}
 \label{definit}
\end{eqnarray}
In the course of time $W(x,t|x_0,t_0)$ is subject to 
possible reset events to $x=0$  or jumps (L\'evy flights). 
Resets are independent from flights and occurring in 
time according to Poisson statistics with  an average expectation time for the 
occurrence of  the event given by $r^{-1}$. {Note that for the purpose of 
analysis, we have untied the initial and resetting positions. We denote the 
former as $x_0$ and keep the latter at the origin.}
The overall process is time homogeneous, 
i.e. $W(x,t|x_0,t_0)=W(x,t-t_0|x_0,0)\equiv W(x,t-t_0|x_0)$, 
so that the propagator satisfies the equation (for the derivation and a detailed discussion see Ref.\cite{gupta2014fluctuating}):
\begin{eqnarray}
W(x,t|x_0)=e^{-rt}W_0(x,t|x_0)+\nonumber \\
+ \int_0^td\tau e^{-r\tau} r W_0(x,\tau|0)
\label{total}
\end{eqnarray}

The first term on the RHS of the above renewal equation represents the survival 
of the probability mass without resetting events, 
whereas the second term describes the evolution after the last reset. 
The function $W_0(x,\tau|x_0,t_0)$ denotes the 
probability density function (PDF) of the process when the resetting mechanism 
is switched off. In this case the random walk propagator fulfills equation 
\begin{eqnarray}
W_0(x,t|x_0)=\delta(x-x_0)\left [1-\int^t_0\Phi(\tau)d\tau) \right]  \nonumber\\
+\int_0^t\Phi(t-t')\int^{+\infty}_{-\infty}
p(x-x')W_0(x',t'|x_0)dx'dt',\nonumber\\
\label{integral}
\end{eqnarray}
where $\Phi(t)$ is the waiting time PDF, independent of the jump-length PDF 
$p(x-x')$.  In the Fourier-Laplace space 
$$W(k,s)\equiv {\cal{F}}[{\cal{L}}[W(x,t);t\rightarrow s];x\rightarrow k],$$ 
the integral Eq.(\ref{integral}) takes the form of
\begin{eqnarray}
W_0(k,s|x_0)=\frac{1-{\Phi}(s)}{s}\frac{1}{1-{\Phi}(s){p}(k)},
\end{eqnarray}
where ${\cal{L}}[f(t)]\equiv \int^{\infty}_0\exp(-st)f(t)dt$. 
We further assume that $\Phi(t)$ has a well defined mean value, 
$\tau_0=\int_0^{\infty}t\Phi(t)dt$, and $p(x)$ is the PDF of the 
L\'evy stable form, so that its characteristic function reads
\begin{eqnarray}
{\cal{F}}[p(x)]=\exp\left[-\sigma^{\alpha}|k|^{\alpha}\right],
\end{eqnarray}
with the stability index $0<\alpha\leq2$. 
The resulting process is Markovian, with the variance diverging 
for $\alpha<2$ and fractional moments \cite{Klagesbook} scaling like:
\begin{eqnarray}
\left<|x(t)|^q\right>\propto (Dt)^{q/\alpha},
\end{eqnarray}
where $D=\sigma^{\alpha}/\tau_0$. 
The asymptotic behavior of $W_0(k,s|x_0,0)$ can be deduced by taking the 
limit $k\rightarrow 0$ and $s\rightarrow 0$ which implies:
\begin{eqnarray}
\Phi(s)\approx 1-s\tau_0+...,\;\;\;p(k)\approx 1-D|k|^{\alpha}.
\end{eqnarray}
After proper rescaling of the waiting times and jumps \cite{Klagesbook}, 
the diffusion limit of the integral Eq.(\ref{integral}) is obtained in 
the form of a space fractional Fokker-Planck equation (FFPE) 
\begin{eqnarray}
\frac{\partial}{\partial t}W_0(x,t|x_0)=
D\frac{\partial^{\alpha}}{\partial |x|^{\alpha}}W_0(x,t|x_0),
\end{eqnarray}
with $ \frac{\partial^{\alpha}}{\partial |x|^{\alpha}}$ denoting the 
symmetric Riesz space fractional derivative which represents an 
integro-differential operator defined as \cite{saichev1997fractional,chechkin2002stationary}:
\begin{eqnarray}
\frac{\partial^{\alpha}}{\partial |x|^{\alpha}}f(x)=
\frac{-1}{2\cos(\pi\alpha/2)\Gamma(2-\alpha)}\times \nonumber \\
\times \frac{\partial^{2}}{\partial x^{2}}
\int^{\infty}_{-\infty}\frac{f(x')}{|x-x'|^{\alpha-1}}dx',
\end{eqnarray}
which has a particularly simple form in the Fourier space
\begin{equation}
{\cal F }\left[\frac{\partial^{\alpha}}{\partial |x|^{\alpha}}f(x)\right] = 
-|k|^{\alpha} {\cal F}[f(x)].
\end{equation}
The total propagator of the process $W(x,t|x_0)$ can then be obtained 
from Eq.(\ref{total}). In the Laplace domain this equation has the form
\begin{eqnarray}
W(x,s|x_0)=W_0(x,s+r|x_0)+\frac{r}{s}W_0(x,s+r|0).
\label{WfromW0}
\end{eqnarray}

In the case of L\'evy flights 
$W_0(k,s|x_0)=\frac{e^{i k x_0}}{D|k|^{\alpha}+s}$.
Hence $W(k,s|x_0)$ is given by 
\begin{equation}
W(k,s|x_0)=\frac{e^{i k x_0}+\frac{r}{s}}{D |k|^{\alpha}+s+r}
\label{Propagator}
\end{equation}
and obeys the differential equation
\begin{eqnarray}
s {W}(k,s|x_0) - e^{i k x_0}=\nonumber \\
=-D |k|^{\alpha} W(k,s|x_0)-r W(k,s|x_0) + \frac{r}{s}.
\end{eqnarray}
The inverse transformationgives the FFPE 
describing the evolution of the total probability distribution:
\begin{eqnarray}
\frac{\partial}{\partial t} W(x,t|x_0)= 
D\frac{\partial^{\alpha}}{\partial |x|^{\alpha}}W(x,t|x_0) - \nonumber \\
- r W(x,t|x_0)+r \delta(x),
\label{eq4propagator}
\end{eqnarray}
with initial condition $W(x,0|x_0)=\delta(x-x_0)$. 
Equation (\ref{eq4propagator}) is analogous to the 
Fokker-Planck equation defining 
a model of diffusion with stochastic resetting \cite{majumdar2011resetting}. 
The difference lies in the fact that instead of a second order spatial 
derivative, characteristic of normal (Gaussian) diffusion, we are dealing now 
with a non-local fractional derivative, which describes L\'evy flights. 
Note that the model analyzed in this paper includes 
the other one as a special case, for $\alpha=2$.

Having calculated the propagator in the Fourier-Laplace space, 
it is straightforward to obtain a characteristic function of the stationary 
distribution. For the sake of simplicity, we also introduce a length 
scale $\lambda^{\alpha}\equiv \frac{D}{r}$. 
By definition, the stationary PDF can be then derived from the relation
\begin{eqnarray}
{p}_s(k;\lambda,\alpha)\equiv\lim\limits_{s\to 0}s W(k,s|x_0) = \nonumber \\
= r W_0(k,s=r|0) = \frac{1}{1+|\lambda k|^{\alpha}}.
\label{Linnik}
\end{eqnarray}
The resulting function, Eq.(\ref{Linnik}), is known as the Linnik distribution 
\cite{linnik1953linear,kozubowski1999univariate}, 
which is a special case of the family 
of geometric stable PDFs, approximating a distribution of normalized sums of 
{\it i.i.d } random variables
\begin{eqnarray}
S_N=\sum_i^Nx_i,
\end{eqnarray}
where the number of terms $N$ is sampled from a geometric distribution, 
i.e. $P(N=k)=(1-p)^{k-1}p$. Summation of that type has been used, among 
others, in modeling energy release of earthquakes, water discharge 
over a dam during a flood, or avalanche dynamics \cite{Pisarenko}.
The Linnik PDF can be expressed in terms of elementary functions only 
for $\alpha=2$, in which case it becomes a well-known Laplace distribution:
\begin{equation}
p_s(x;\lambda,2)=\frac{1}{2 \lambda} e^{-\frac{|x|}{\lambda} },
\label{Laplace}
\end{equation}
with a zero mean and a variance $\mathrm{Var}[x^2]=2 \lambda^2$.
For $\alpha=1$ the closed-form expression for the corresponding Linnik PDF can 
be obtained (cf. Appendix \ref{app-linnik}) in terms of special functions 
$\mathrm{Si}(x)\equiv\int\limits_{0}^{x} \frac{\sin{t} d t}{t}$ and 
$\mathrm{Ci}(x)\equiv-\int\limits_{x}^{\infty} \frac{\cos{t} d t} {t}$, 
and in a scaled form reads:
\begin{equation}
\lambda p_s(\lambda x;1,1)=
\left(\frac{1}{2}-\frac{\mathrm{Si}(x)}{\pi}\right)\sin{|x|}-
\frac{1}{\pi}\mathrm{Ci}(|x|) \cos{x}.
\label{Linnik1}
\end{equation}
When passing to the analysis of the first arrival times in a subsequent 
Section, we note here that the result \eqref{Linnik} has been obtained 
earlier in Ref.\cite{anderson1993linnik} for a discrete time counterpart of the 
resetting model.

\section{The problem of  the first arrival time}

For the stochastic process defined by Eqs.(\ref{definit},\ref{total}), 
a question of interest is in the estimation of the waiting time before the 
first event of a magnitude greater than a given threshold is observed. 
However, as has been discussed elsewhere 
\cite{chechkin2003first,Dybiec2006,Dybiec2009}, the superdiffusive nature of 
L\'evy flights strongly influences the statistics of first passage times over 
the threshold. In particular, due to long-range L\'evy jumps 
occurring with an appreciable probability, the trajectory of the process may 
cross the threshold numerous times without actually hitting it. In consequence, 
the statistics of first arrival times at a predefined barrier is different from 
the statistics of first passages over it. 

Following Refs.\cite{chechkin2003first,redner2001guide}, 
we introduce the first arrival time PDF $p_{fa}(t,x)$, 
which describes the distribution of times $T_{fa}$
in terms of the integral equation for the propagator $W(x,t|0,0)$:
\begin{equation}
W(x,t|0,0)=\int\limits_0^t d \tau \mbox{ }p_{fa}(\tau,x) W(x,t|x,\tau).
\label{fa}
\end{equation}
The above formula can be easily interpreted: it simply states that the process 
which at time $t$ finishes up at $x$, has had to get to that point for 
the first time at some time $\tau\in(0,t)$. 
After that it could move freely until at time $t$, 
it came back to the very same point. The assumption of 
time-homogeneity ($W(x,t|x,\tau)=W(x,t-\tau|x,0)$) explains a convolution 
operator on the RHS of  Eq.(\ref{fa}). 
The function $p_{fa}$ is a probability density function of its first argument. 
The second argument denotes 
that the first arrival to a position $x$ is evaluated. For readability, 
we skip $D$, $r$ and $\alpha$ in the parameter list. From now on, we also 
assume that the initial and reset positions coincide. 

By transforming Eq.(\ref{fa}) into the Laplace space 
a simple algebraic relation is obtained:
\begin{equation}
{W}(x,s|0)={p}_{fa}(s,x){W}(x,s|x).
\label{WandPfa}
\end{equation}
It is important to notice that ${W}(x,s|x)\neq {W}(0,s|0)$, 
as the resetting mechanism introduces space inhomogeneity. 
Our aim is to derive a formula for the mean first arrival time (MFAT) 
which can be obtained from ${p}_{fa}(s,x)$ as follows:
\begin{equation}
\langle T_{fa}(x) \rangle = 
-\frac{\partial}{\partial s} {p}_{fa}(s,x)|_{s=0} 
=\frac{1-{p}_{fa}(s,x)}{s}|_{s=0}.
\label{TzPfa}
\end{equation}
We proceed by inserting the propagator, Eq.(\ref{WfromW0}), 
and the algebraic relation between the propagator and $p_{fa}(s)$, 
Eq.(\ref{WandPfa}), into the formula for MFAT, Eq.(\ref{TzPfa}). 
After straightforward algebraic manipulations we arrive at: 
\begin{eqnarray}
\langle T_{fa}(x) \rangle=
\frac{1}{r}\left ( \frac{{W}_0(x,s=r|x)}{{W}_0(x,s=r|0)}-1\right )=
\nonumber \\
=\frac{1}{r}\left (\frac{p_s(0;\lambda,\alpha)}{p_s(x;\lambda,\alpha)}-1\right).
\label{TfromPropagator}
\end{eqnarray}
Note that for simplicity we use a shortened notation 
$W_0(x,t)\equiv W_0(x,t|0,0)$.
Eq.(\ref{TfromPropagator}) shows that the MFAT can be expressed either 
in terms of the Laplace transform of the propagator of the 
standard L\'evy $\alpha$-stable process without resetting, 
or in terms of the stationary PDF of the process with the resetting mechanism 
switched on. This result is very general, 
since in the derivation no particular form of $W_0(x,t|x_0)$ has been assumed.

We further focus on the special case of L\'evy flights. 
In general, the propagator of a L\'evy stable process cannot be expressed in 
terms of an elementary function of $x$. 
Representations in terms of the Fox functions \cite{metzler2000random} 
and in terms of the generalized hypergeometric functions \cite{gorska2011levy}
are known, but they are not useful in our case. 
We can, however, calculate $W_0(x,s=r|x)$ and deduce from its form the 
range of the stability parameter $\alpha$ that guarantees finiteness of the 
evaluated MFAT:
\begin{eqnarray}
{W}_0(x,s|x)=\frac{1}{2\pi}\int\limits_{R}d k 
\int\limits_{0}^{\infty} d t e^{-s t} e^{- D |k|^{\alpha} t}=\nonumber \\
=\frac{\Gamma(\frac{1}{\alpha}) \Gamma(1-\frac{1}{\alpha})}
{\pi \alpha D^{\frac{1}{\alpha}} s^{1-\frac{1}{\alpha}}} =
\frac{1}{\alpha \sin{\frac{\pi}{\alpha}} 
D^{\frac{1}{\alpha}} s^{1-\frac{1}{\alpha}}}.
\label{stala}
\end{eqnarray}
For any $x\neq0$, the propagator ${W}_0(x,r|0)$ is finite, since it is an 
integral of an oscillating function with an amplitude decreasing to zero, 
and can be rewritten as an alternating series. 
We therefore conclude from Eq.(\ref{stala}) that the MFAT diverges for 
$\alpha \leq 1$ and remains finite for $1<\alpha \leq 2$. 
That apparent finiteness of the MFAT in case of L\'evy flights is rather 
surprising, taking into account the discontinuous character of superdiffusive 
trajectories and thus the possibility of overshooting 
(i.e. jumping over the target).

\subsection{Asymptotic behavior}

The average $\langle T_{fa}(x) \rangle$ cannot be expressed in terms of 
elementary functions for arbitrary $\alpha$. Nevertheless, we can learn 
something about its behavior for large and small distances  $x$ to a target. 
By taking a well-known expression for the asymptotic expansion of 
$\alpha$-stable distributions \cite{metzler2000random} and transforming it to 
the Laplace space, or otherwise, directly expanding:
\begin{equation}
\frac{1}{D|k|^{\alpha}+s}=
\sum\limits_{n=0}^{\infty}\frac{(-D|k|^{\alpha})^{n}}{s^{n+1} }
\end{equation}
and transforming this back from the Fourier space term by term, 
we obtain an asymptotic expansion of the propagator $W_0$ in the Laplace 
space (see also \cite{kotz1995analytic,kotz2001laplace} 
for more formal derivations):
\begin{equation}
W_0(x,s|0,0)=
\frac{1}{\pi}\sum\limits_{n=1}^{\infty}(-1)^{n+1} 
\sin(\frac{\pi}{2} n \alpha) \frac{D^{n} 
\Gamma(n \alpha +1)}{s^{n+1} x^{n \alpha +1}}.
\label{asymptoticexpansion}
\end{equation}
This expression is correct for $\alpha \in (1,2)$. 
For $\alpha=2$ we don't need the asymptotic expansion since in this case we 
have a closed-form expression:
\begin{equation}
W_0(x,s|0,0)=\frac{1}{2 \sqrt{D s} }e^{-|x|\sqrt{\frac{s}{D}}}
\mbox{ }\mbox{ }\mbox{ }(\mbox{for }\alpha=2).
\label{exactW0}
\end{equation}
One can easily verify that Eqs.(\ref{TfromPropagator},\ref{stala}) 
together with Eq.(\ref{exactW0}) give the same result as the one derived 
in \cite{majumdar2011resetting}. We truncate the series at the first term and 
so obtain the large $x$ behavior of the MFAT:
\begin{equation}
\langle T_{fa}(x) \rangle \propto
\begin{cases}
x^{\alpha+1} &; 1<\alpha <2 \\
e^{x\sqrt{\frac{r}{D}}} &; \alpha=2
\end{cases}
\label{largez}
\end{equation}

We may also expand the MFAT around $x=0$ using the known expansion of  the 
Linnik distribution \cite{kotz1995analytic,kotz2001laplace}. This leads to:
\begin{equation}
\langle T_{fa}(x) \rangle \approx \frac{\alpha \sin{\frac{\pi}{\alpha}}}
{2\sin{\frac{\pi (\alpha-1)}{2}}\Gamma(\alpha)}\frac{1}{r^{\frac{1}{\alpha}} 
D^{1-\frac{1}{\alpha}}}x^{\alpha-1}+O(x^{2\alpha-2}).
\label{smallz}
\end{equation}

\section{Optimization of the MFAT}

Given a distance to a target $x$, one could be tempted to determine the 
optimal search kinetics of this location.
We choose MFAT as an objective function, and minimize it in the space of 
parameters $(r,\alpha)$. We will denote derived parameters of the efficient 
strategy as $r^*(x)$, $\alpha^*(x)$, respectively and the corresponding 
optimal MFAT as $T^*(x)$.

\subsection{Fair comparison}
Since we want to compare L\'evy flights with different stability indices 
$\alpha$, it is important to carefully choose the parametrization of the 
family of jump distributions. One commonly used is $\phi(k)=e^{-|k|^\alpha}$ 
which in our case means fixing $D=1$ for every $\alpha$. 
Alas, this choice is very arbitrary and based on simplicity of 
a characteristic function for symmetric stable distributions.  
As an alternative option, we propose here a straightforward and consistent 
approach based on fractional moments. 
Let us define a random variable $\xi_{\alpha}$ to be a position of the 
process without resetting at time $t=1$ (this fixes the time unit). 
The $p$-th fractional moment may be expressed as
\begin{equation}
\lambda_0^p = \langle|\xi_{\alpha}|^p\rangle=
D^{\frac{p}{\alpha}} f(\alpha,p),
\end{equation}
where the condition $p<\alpha$ has to be satisfied in order for 
the fractional moment to be finite. 
Function $f(\alpha,p)$ is known and reads \cite{shao1993signal}
\begin{equation}
f(\alpha,p)=\frac{2^{p+1}\Gamma(\frac{p+1}{2})\Gamma(-\frac{p}{\alpha})}
{\alpha \sqrt{\pi} \Gamma(-\frac{p}{2})}.
\end{equation}
We want to keep $\lambda_0$ constant (e.g. $\lambda=1$) so our $D$ will 
depend on $\alpha$ and $p$. The most natural choice of $p$ in our case is 
$p=1$ since it does not exclude any solution 
(in line with findings of Section III, we refer to cases with $\alpha>1$ 
assuring finiteness of MFAT) and it induces an $L_1$ norm that is commonly used 
in many applications. This choice leads to the expression:
  \begin{equation}
  D(\alpha)=\left( \frac{\pi}{2 \Gamma(1-\frac{1}{\alpha})}\right)^{\alpha}.
  \end{equation}
In the following we will refer to this method of comparison, based on the 
choice $p=1$, as  the ``fair comparison''. 
This is in contrast to the ``naive comparison'' 
based on the simplicity of characteristic function $(D=1)$.

\subsection{Asymptotic analysis}

From the asymptotic behavior of the MFAT several conclusions may be drawn: 
The prefactor in Eq.(\ref{smallz}) is bounded for $\alpha\in(1,2]$. 
Consequently, for  given non-zero $r$ and $D$ it is always possible to find $x$ 
small enough, so that $\alpha=2$ minimizes the MFAT. In other words, 
Brownian motion is expected to be the optimal strategy at small distances to 
the target. In contrast, as it can be inferred from the asymptotic behavior, 
Eq.(\ref{largez}), for large enough distances the MFAT increases with $x$ much 
faster for $\alpha=2$ than for $\alpha<2$. In this case, the L\'evy motion with 
$\alpha<2$ minimizes the MFAT, thus indicating a more efficient kinetics of 
space exploration to detect a target.

 \subsection{Random distribution of target sites} 
 \label{sec-random-dist}
 
 In many natural scenarios, living organisms navigate to unpredictable or 
 randomly distributed resources. 
 In other words, positions of the "target" is not precisely known. 
 How is the kinetics of random search with resetting affected 
 by the location of targets in an unknown environment? In order to address this 
 point, we further explore the MFAT under the constraint that the searcher 
 knows only the mean (expected) distance to the target.
 Accordingly, instead of a fixed $x$ in the evaluation of the MFAT, 
 we use the PDF that satisfies the maximum entropy principle, 
 i.e. a Laplace distribution $p(x)$ of target positions is assumed. 
 The MFAT in this more general setting 
 can be calculated by averaging over possible distances:
 \begin{eqnarray}
 \langle T_{fa}(\lambda_t)\rangle=
 \int\limits_{-\infty}^{\infty}d x \langle T_{fa}(x)\rangle p(x)=\nonumber \\
 =\frac{1}{2 \lambda_t} \int\limits_{-\infty}^{\infty}d x \langle T_{fa}(x)
 \rangle e^{-\frac{|x|}{\lambda_t}}.
 \label{mfat-dist}
 \end{eqnarray}
 Even though $\langle T_{fa}(\lambda_t)\rangle$ is a different  function from 
 $\langle T_{fa}(x)\rangle$, for readability we keep the same symbol for the 
 MFAT averaged over the distribution of targets and denote that by use of a 
 different argument, only.
 
As explained in the following example, such averaging over random distances 
to a target leads to modification of the MFAT and becomes crucial for the 
optimal strategy planning. Let us assume Brownian diffusion $\alpha=2$ with 
the Laplace PDF of target positions characterized by the mean distance to the 
target $\langle|x|\rangle = \lambda_t$. 
In that case the MFAT is given by the formula:
 \begin{equation}
 \langle T_{fa}(\lambda_t)\rangle=\frac{1}{r}
 \frac{1}{\frac{\lambda}{\lambda_t}-1},
 \label{average}
 \end{equation}
where $\lambda=\sqrt{\frac{D}{r}}$. Clearly, the MFAT is finite for 
$\lambda \geq \lambda_t$ and optimization of Eq.(\ref{average}) yields the 
value of the resetting frequency $r^*_2(\lambda_t)=\frac{D}{4 \lambda_t^2}$. 
If a searcher does not know the distribution of target locations, but was able 
to estimate via several measurements the mean distance to the target, 
$\langle|x|\rangle\approx \lambda_t$, 
he might be prompted to use that fixed position for further optimization of 
the MFAT, $\langle T_{fa}(x=\lambda_t)\rangle$. The derived optimal $r^*$, 
see Eq.(\ref{opt_Brownian}), when applied to the system with Laplace 
distributed distance-to-target, would then lead to an infinite MFAT. 
This apparent inconsistency demonstrates that for the proper minimization of 
arrival times, the actual form of distance-to-target distribution 
$p(x)$ is indispensable.
 
It can be easily shown that for heavy-tailed distance-to-target distributions, 
the Brownian strategy always gives an infinite MFAT.  
In contrast, strategies with 
L\'evy-distributed jumps ($\alpha<2$) may provide efficient algorithms for 
searching, for which the MFAT remains finite 
as long as the $p(x)$ distribution is 
characterized by a finite variance. A simple example illustrating this case is 
optimization of the MFAT given by Eq.(\ref{mfat-dist}) with the 
Student's t-distribution of distances to the target:
\begin{equation}
p(x)=\frac{\Gamma\left(\frac{\nu+1}{2}\right)}{\sqrt{\nu \pi} 
\Gamma\left(\frac{\nu}{2}\right) \lambda_t} 
\left(1+\frac{x^2}{\nu \lambda_t^2} \right)^{-\frac{\nu+1}{2}}.
\label{studentt}
\end{equation}
In this case the integral in Eq.(\ref{mfat-dist}) is convergent iff condition 
$\alpha < \nu - 1$ holds.
Numerical integration of Eq.(\ref{mfat-dist}) for $\nu=2.7$ and $\nu=4$ leads 
to MFAT functions displayed in 
Figs.\ref{fig:optimal_student4} and \ref{fig:optimal_student27}. 

 \subsection{Scaling}
Optimal parameters $r^*(x)$, $\alpha^*(x)$ and optimal MFAT $T^*(x)$ 
depend on $x$ and $D$. For the sake of simplicity, from now on we fix $D$. 
It will be useful to take advantage of dimensional analysis to calculate the
scaling behavior of the optimal $r^*$ and MFAT for a given $\alpha$. 
Let $r^*_{\alpha}(x)$ and $T^*_{\alpha}(x)$ be the optimal $r$ and the optimal 
MFAT for fixed $x$, $\alpha$ and $D$. 
Up to an arbitrary multiplicative constant, the only combination of $x$ and $D$ 
that has the dimension of time is $t=\frac{x^\alpha}{D}$. 
This leads to the following scaling equations:
\begin{equation}
\begin{cases}
T^*_{\alpha}(x)=T^*_{\alpha}(1) x^{\alpha}\\
r^*_{\alpha}(x)=\frac{r^*_{\alpha}(1)}{x^{\alpha}}.
\end{cases}
\label{scaling}
\end{equation}
One easily verifies that these equations hold, by calculating the derivative of 
the MFAT (Eq.\ref{TfromPropagator}) with respect to $r$, comparing it to 0, and 
rewriting the corresponding equation such that it contains only a function of 
$r  x^{\alpha}$. Scaling equations (\ref{scaling}) also imply similar relations 
to be fulfilled by  $T^*(\lambda_t)$:
 \begin{equation}
\begin{cases}
T^*_{\alpha}(\lambda_t)=T^*_{\alpha}(1) \lambda_t^{\alpha}\\
r^*_{\alpha}(\lambda_t)=\frac{r^*_{\alpha}(1)}{\lambda_t^{\alpha}}.
\end{cases}
\label{scaling_lambda}
\end{equation}
The above relations are used in a numerical algorithm for optimization, 
as explained in details in the Appendix \ref{app-numerical-scheme}.

\subsection{Results}

\begin{figure}
\begin{center}
\includegraphics[width = 1.0\linewidth]{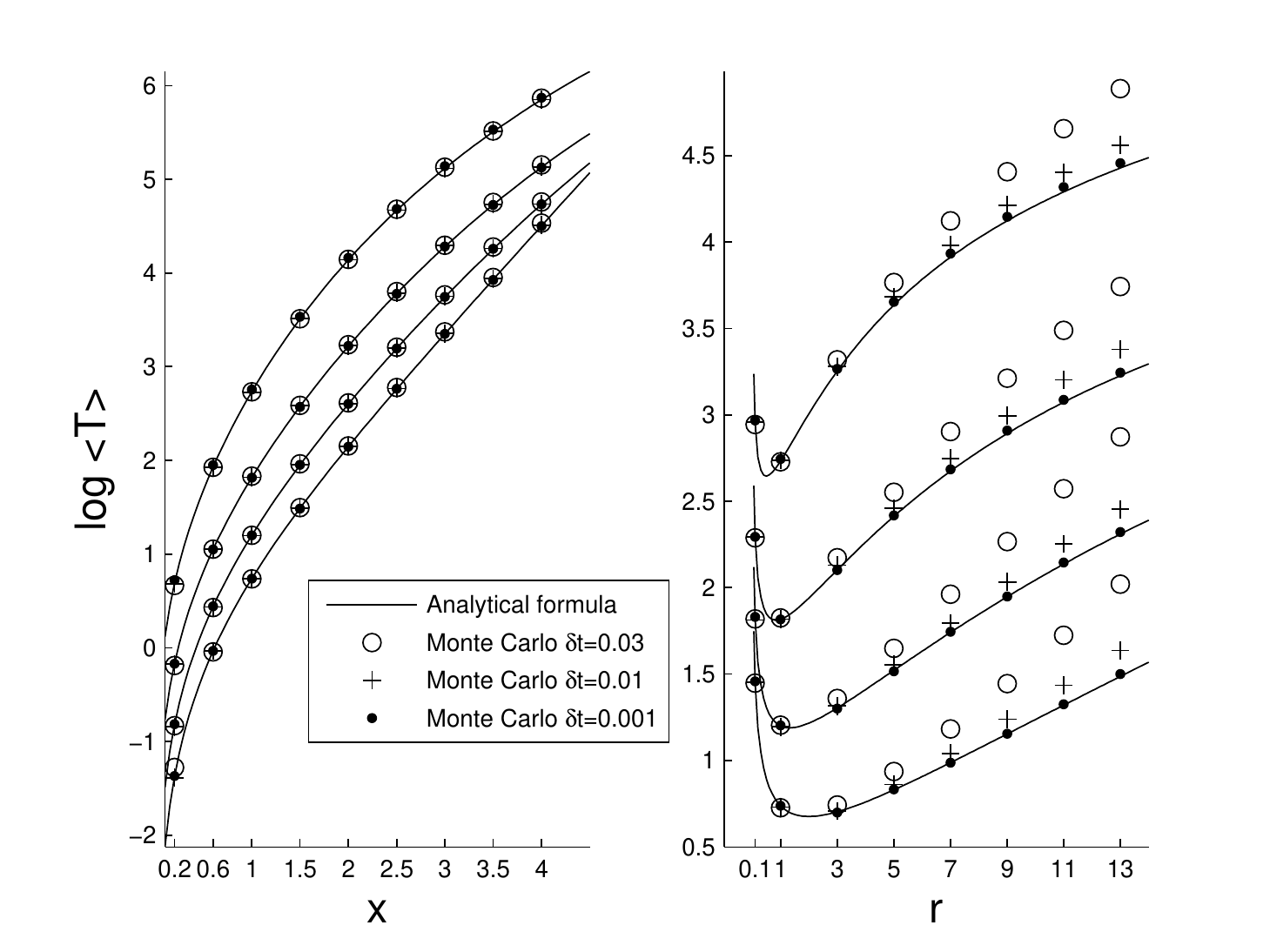}
\caption{{Comparison between MFATs obtained by numerical integration of 
the analytical formula (\ref{TfromPropagator}) (lines) and by averaging over 
$N=10^5$ realizations of a simulated process. Different lines 
(from the top to the bottom) correspond to $\alpha=(1.4, 1.6, 1.8, 2)$. 
For the sake of simulation not only time has to be discretized ($\delta t$), 
but also a finite target size is needed. For each $\alpha$ the target size is 
chosen separately to match the analytical result at $x=1$, $r=1$. 
The same target size is further used across different values of $x$ and $r$. 
Estimated error bars are smaller than the markers used in the plots and hence 
have not been displayed.}}
\label{fig:mc}
\end{center}
\end{figure}

\begin{figure}
\begin{center}
\includegraphics[width = 1.0\linewidth]{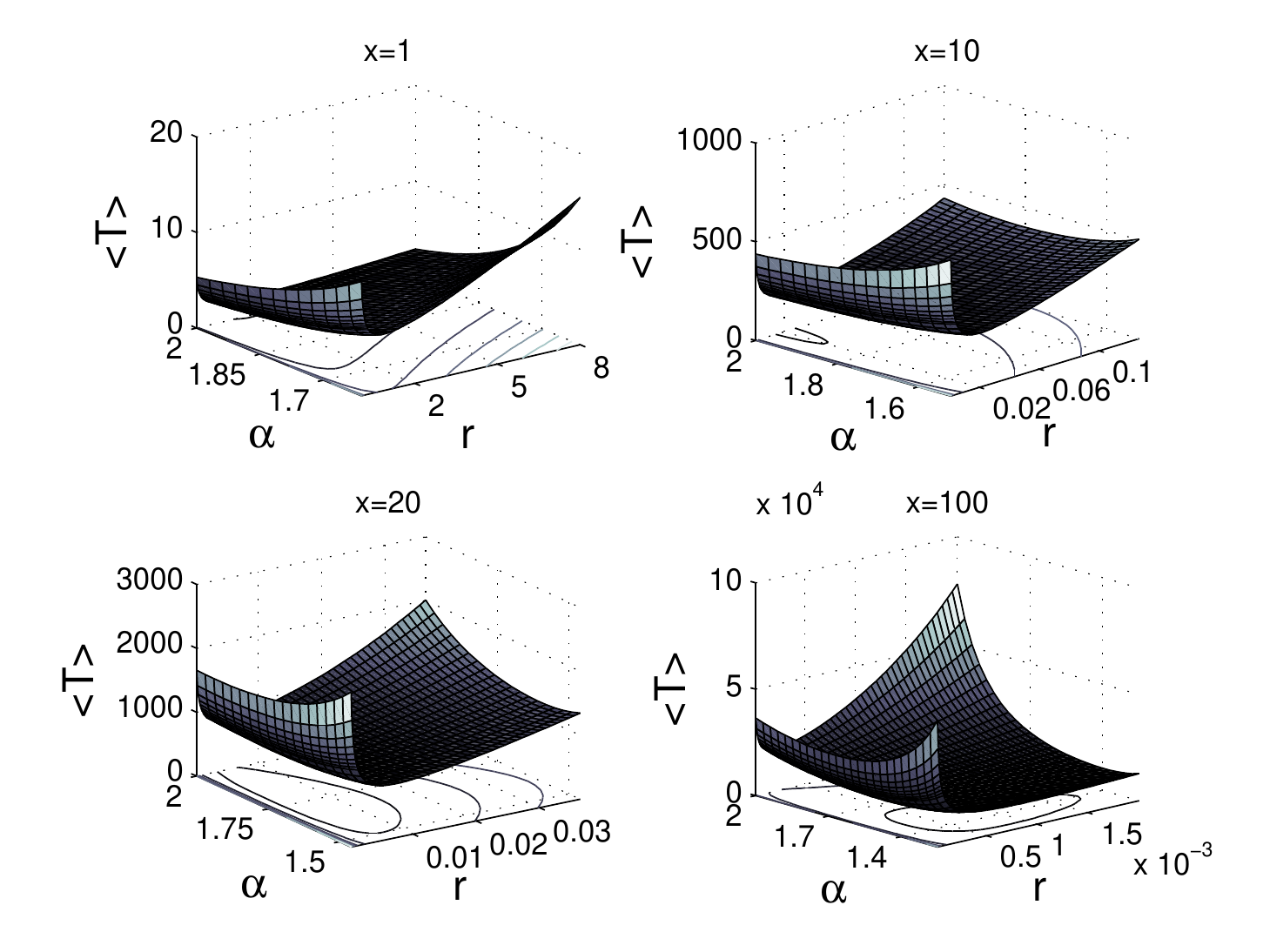}
\caption{{The MFAT as a function of the parameters $(\alpha, r)$ for different 
values of the distance to an immobile target $x=(1,10,20,100)$. 
Contour plots beneath the surfaces help to guide an eye towards the minimum.}}
\label{fig:3d}
\end{center}
\end{figure}

A comparison between analytical prediction, 
Eq.(\ref{TfromPropagator}), and numerical stochastic simulations has been 
performed and the results are displayed in Fig.\ref{fig:mc} demonstrating 
a perfect agreement between both approaches. 
Additionally, Fig.\ref{fig:3d} presents the analytically derived 
MFAT functions in 2-dim $(\alpha,r)$ parameter space.

The MFAT diverges as $r\rightarrow 0$ and $r\rightarrow\infty$ 
(cf. Figs. \ref{fig:mc},\ref{fig:3d}). Accordingly, a minimum of the MFAT with 
respect to $r$ can be found in the interval $[0,\infty)$ and its position 
depends on the stability index $\alpha$ characterizing 
underlying diffusive process. 

For small $x$ MFAT values are systematically higher for non-Gaussian diffusion 
($\alpha<2$) than for the Gaussian case and the same resetting rates. 
Also, as displayed in Fig.\ref{fig:mc}, the MFAT has a more pronounced, deeper 
minimum in function of $r$ for L\'evy diffusion with heavier tails 
(i.e. lower $\alpha$'s), which suggests that the Gaussian strategy 
is more robust to variations of $r$.
This is, however, no longer true for large $x$ (cf. Fig.\ref{fig:3d}). 
In that case MFAT values for $\alpha=2$ are higher than that for 
$\alpha<2$ and the same $r$, at least in the vicinity 
of the optimal $r^*_{\alpha}$. 
Moreover, in this limit L\'evy flights become more resilient to changes in $r$, 
especially in the range $r\geq r^*_{\alpha}$.

Results displayed in Fig.\ref{fig:3d} have been further analyzed to derive 
minimal values of the MFAT with respect to a pair of parameters $(\alpha, r)$ 
for different values of a distance to a target, $x$. 
Consecutive Fig.\ref{fig:optimal_x} and Fig.\ref{fig:optimal_lambda} show 
outcomes of the optimization procedure described in 
Appendix~\ref{app-numerical-scheme} for the cases 
of the immobile target located at a distance $x$, and the target with position 
described by Laplace distribution with an average distance to a target 
$\lambda_t$, respectively.

No qualitative difference in the derived optimal MFAT values 
has been found between the naive and the fair comparison. 
We therefore present results of the numerical optimization of 
$\langle T_{fa}(x)\rangle$ and $\langle T_{fa}(\lambda_t)\rangle$ 
for the fair comparison only. 

As expected, for small $x$ ($\lambda_t$) Gaussian 
diffusive motion ($\alpha^*=2$) is the optimal searching strategy. 
With growing distance to a target $x$ (or $\lambda_t$) the minimum of 
$\langle T_{fa}\rangle$ becomes shallower, up to some point 
$x^*\approx10.8$ ($\lambda_t^*\approx3.25$), beyond which Gaussian 
diffusion is not efficient anymore and the optimal stability index switches to 
values $\alpha^*<2$. Corresponding values of bifurcation points $x^*$ and 
$\lambda_t^*$ have been obtained by means of a numerical optimization 
procedure and are marked in 
Figs.\ref{fig:optimal_x},\ref{fig:optimal_lambda} with a cross sign. 

The described scenario of the continuous transition between 
the Gaussian and non-Gaussian optimal strategies is qualitatively similar to 
the one investigated in Ref.\cite{palyulin2014levy}. 
In that article yet another variant of a one-dimensional L\'evy flight search 
strategy has been analyzed: 
The optimization of the random search for targets has been performed 
with respect to the average over inverse search times. 
The model has been enriched with a nonzero drift term (representative of an external bias or former experience of the searcher) and no resetting mechanism has been included. 
Despite these differences, their plot of the optimal $\alpha^*$ as a function 
of the initial position $x$ (see Fig. 3, Ref.\cite{palyulin2014levy}) at vanishing drift strength looks very similar to ours findings in Figs. 3 and 4:
There exists a finite region (of relatively small $x$'s) in which the Brownian diffusion is 
the most efficient strategy and the optimal $\alpha^*$ is always larger than $1$. 
It seems that these observations are generic features of the analyzed optimal first arrival times.

We have also investigated the impact of the heavy-tailed distribution 
of distances to the target on the efficiency of the searching. 
The analysis of the the optimal MFAT performed in this case is illustrated in 
Figs.\ref{fig:optimal_student4} and \ref{fig:optimal_student27}. 
Presented plots indicate that the heavy-tailed distribution 
of distance-to-target excludes Gaussian diffusion, $\alpha=2$, 
from the set of possible optimal search strategies. 
Moreover, in line with the analysis of Section \ref{sec-random-dist}, 
for L\'evy flights a condition $\alpha<\nu-1$ has to be met in order to 
perform a successful search with a finite MFAT.

\begin{figure}
\begin{center}
\includegraphics[width = 1\linewidth]{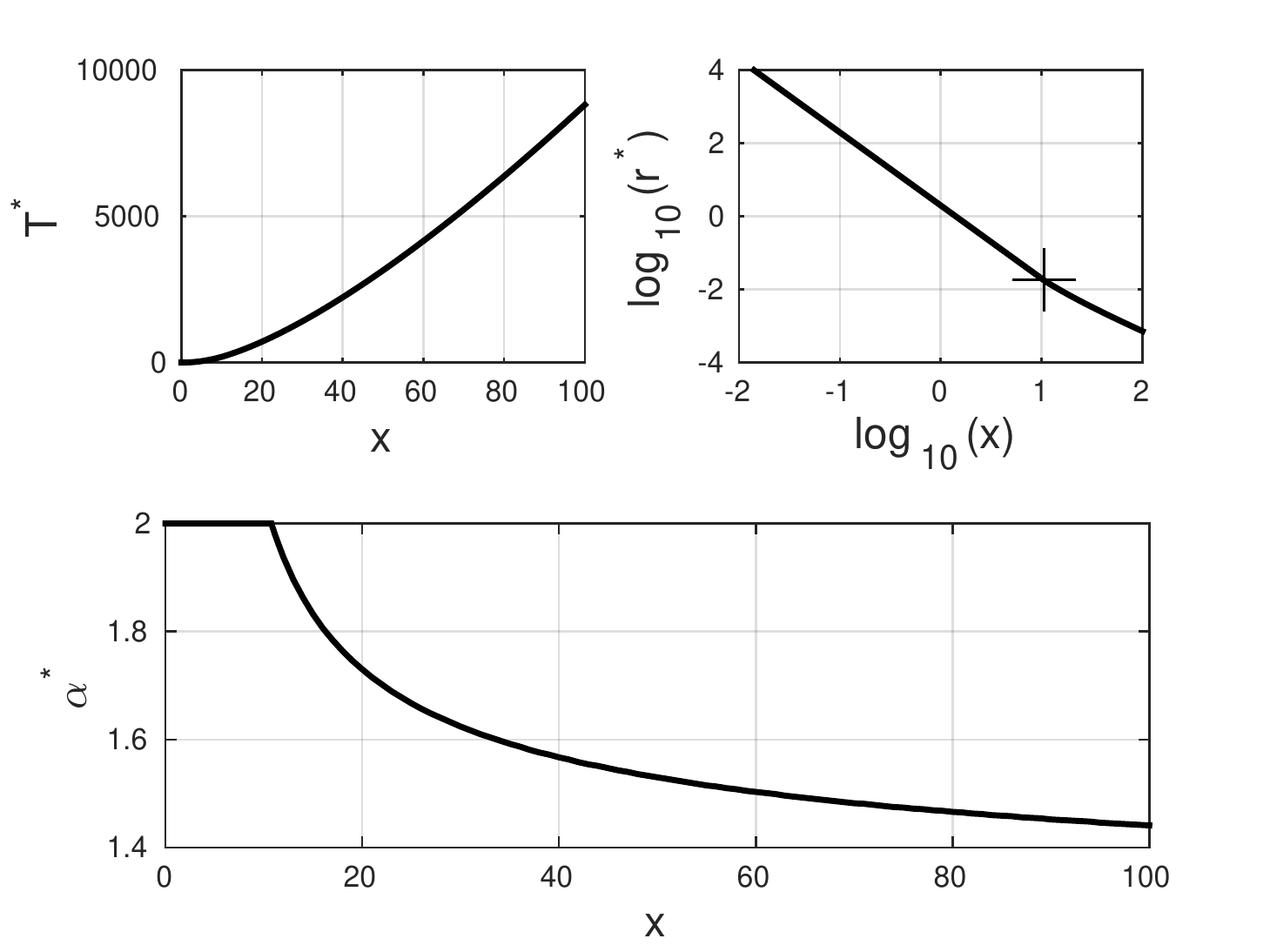}
\caption{Optimal parameters $(\alpha^*,r^*)$ and the MFAT as functions of the 
distance to a target $x$. }
\label{fig:optimal_x}
\end{center}
\end{figure}

\begin{figure}
\begin{center}
\includegraphics[width = 1\linewidth]{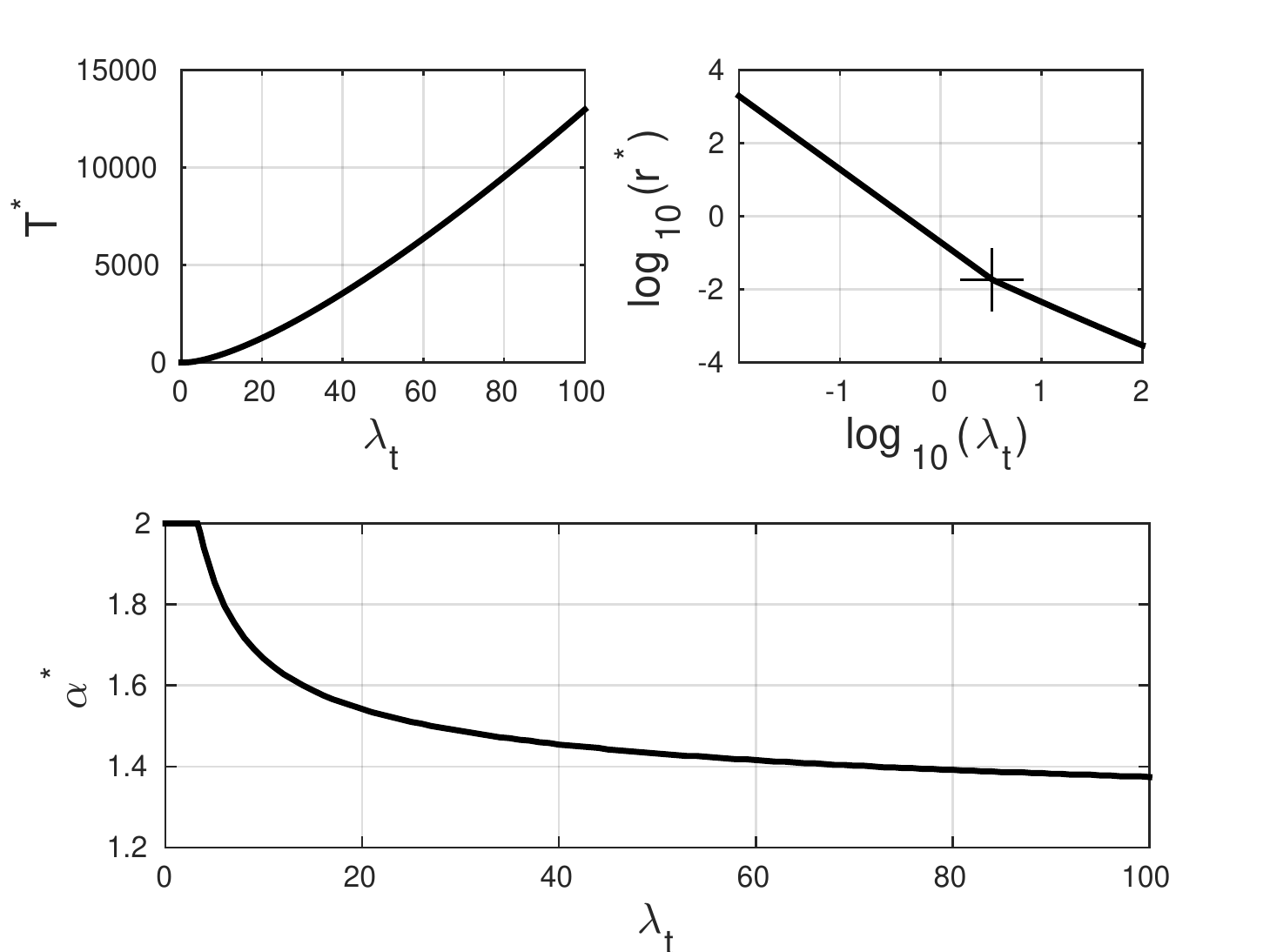}
\caption{Optimal parameters $(\alpha^*,r^*)$ and the MFAT as functions of 
$\lambda_t$. The target position is a random variable with a Laplace PDF 
of distances and an average distance-to-target $\lambda_t$.}
\label{fig:optimal_lambda}
\end{center}
\end{figure}

\begin{figure}
\begin{center}
\includegraphics[width = 1\linewidth]{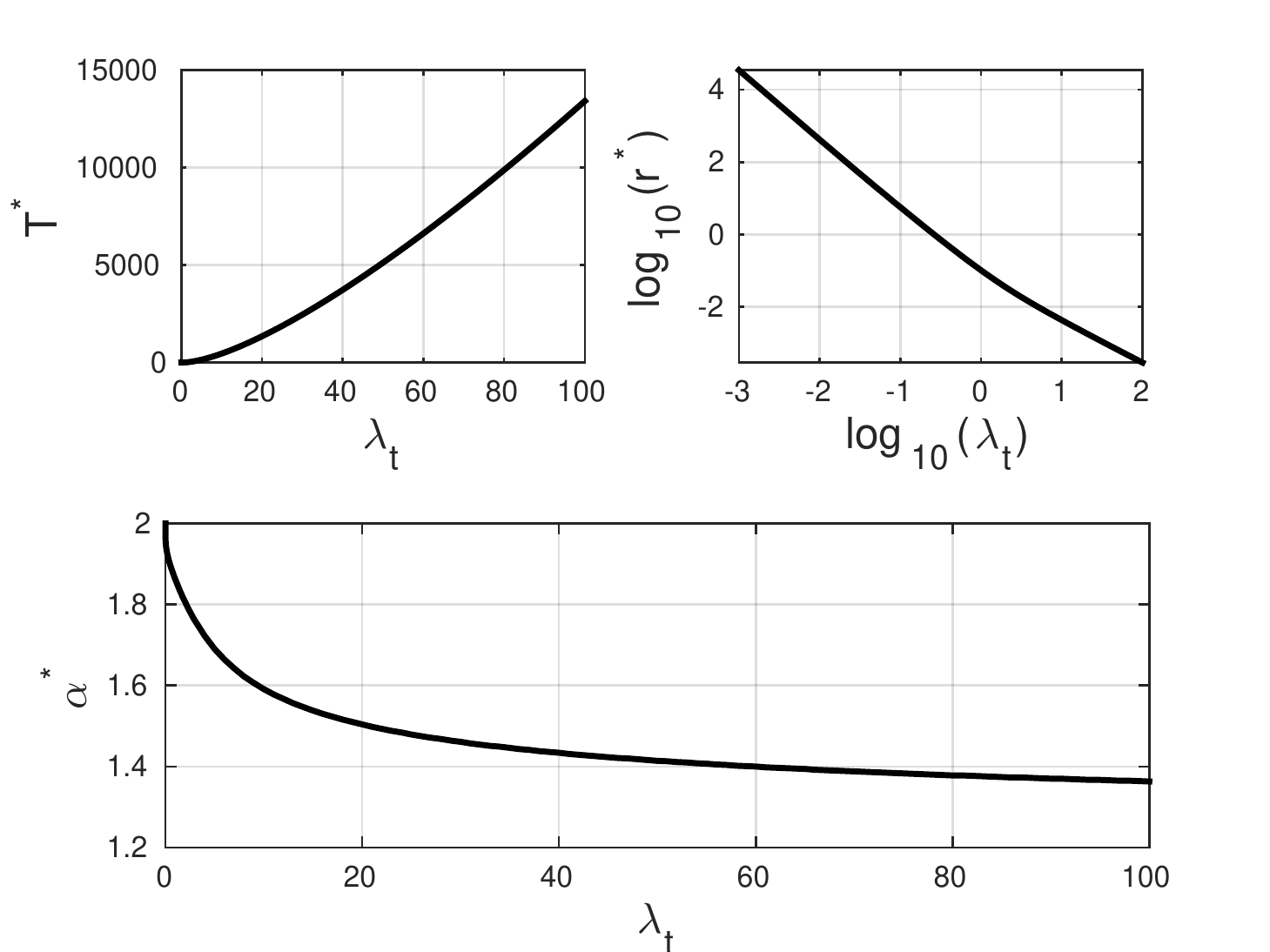}
\caption{Optimal parameters $(\alpha^*,r^*)$ and the MFAT as functions of 
$\lambda_t$. 
The target position is given by a Student's t-distribution, Eq.(\ref{studentt}),
 with $\nu=4$ and an average distance-to-target $\lambda_t$.}
\label{fig:optimal_student4}
\end{center}
\end{figure}

\begin{figure}
\begin{center}
\includegraphics[width = 1\linewidth]{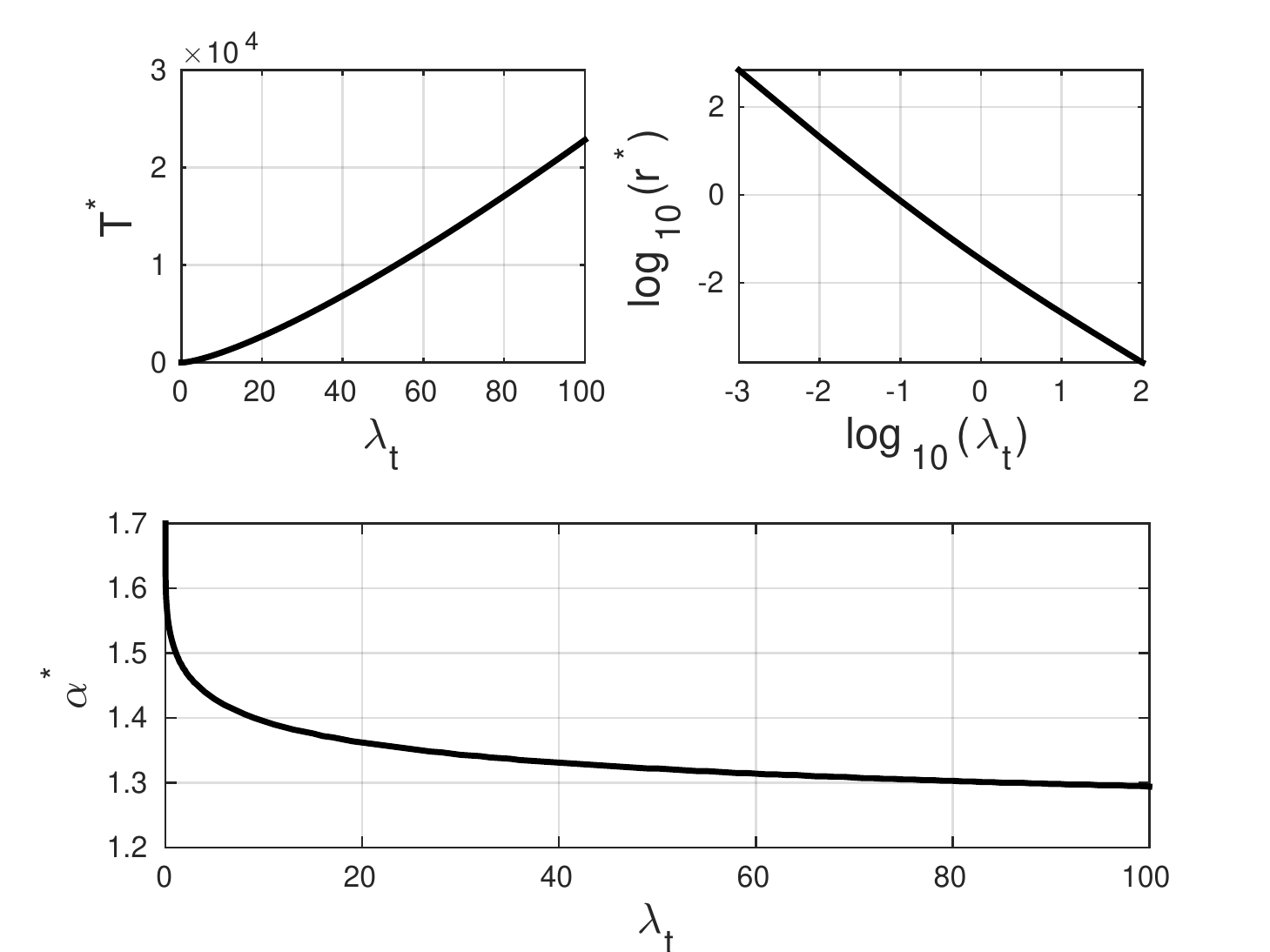}
\caption{Optimal parameters $(\alpha^*,r^*)$ and the MFAT as functions of 
$\lambda_t$. Target position is given by a Student's t-distribution, 
Eq.(\ref{studentt}), with $\nu=2.7$ and an average distance-to-target 
$\lambda_t$.}
\label{fig:optimal_student27}
\end{center}
\end{figure}

\section{Conclusions} 

Not only animal foraging patterns, but also memory retrievals of humans 
\cite{Kello} and fluctuations of their spontaneous activity \cite{Ochab} 
exhibit scaling statistics. The problem devised in this paper models mechanism 
of stochastic resetting, or relaxation of a diffusive searching process to a 
predefined threshold, and as such can be well adapted to many natural 
scenarios of exploration processes such as, e.g. quests for food in a given 
territory \cite{Reynolds2015}, translocation and recruitment of repair proteins 
seeking for a disrupted DNA strand to be repaired \cite{Badrin}, 
optimal computer-aided web search \cite{tong2006fast} or statistics of recall 
periods in retrospective memory \cite{Kello}.

The efficiency of a search may be defined and analyzed by use of different 
measures, like e.g. the number of encounters of searchers and targets per 
unit of time, the mean inverse search time \cite{palyulin2014levy}, 
or the exploration range of space per unit of time. 
Here, we have focused on the efficiency measure expressed 
by the mean time to reach an immobile target, the MFAT. 

The first arrival time statistics has been analyzed for the one-dimensional 
problem with a constant resetting rate $r$. The acts of trajectory relocation 
have been assumed independent from the free (super)diffusive motion 
described by L\'evy jumps with the exponent $0<\alpha\leq 2$. 
Despite the discontinuity of trajectories, typical for L\'evy flights, the MFAT 
remains finite iff $\alpha>1$ with a rich characteristics of optimal (minimal) 
times $T^*(x)$. By use of the designed optimization method (Section IV), 
we have been able to derive the optimal parameters 
$r^*(x)$ and $\alpha^*(x)$ for the range of target positions $x$. 
We have shown that the randomized distribution of targets with some average 
distance to a target results in a severe reduction of distances for which 
Gaussian search remains the optimal strategy. 
Moreover, our analysis of optimal searching times for exponential distribution 
of distances to a target (Section \ref{sec-random-dist}) clearly indicates that 
not only first moment of that distribution but rather its actual form is needed 
for a proper optimization planning: an optimization procedure based solely on 
the information about the average distance to a target would result in the 
optimal $r^*$ leading to an infinite MFAT.
 
Altogether, the proposed optimization scheme and scaling analysis can be 
further exploited, e.g. for two- and three-dimensional searching scenarios. 
Another plausible modification of the proposed procedure could be 
an implementation of L\'evy walks, with coupled space-time distributions, 
or truncated L\'evy flights, penalizing very long jumps.

\section{Acknowledgments}
This project has been supported in part (EGN) by 
National Science Center (ncn.gov.pl), a grant no. DEC-2014/13/B/ST2/020140.
Authors acknowledge many valuable discussions with Martin Bier.
\appendix
\section{Linnik distribution}
\label{app-linnik}
The derivation of the Linnik PDF, expressed in terms of $p_s(x,\lambda,1)$ in 
Eq.(\ref{Linnik1}) proceeds as follows. Since $p_s(x,\lambda,1)$ as a function 
of $x$ is even, without loss of generality,  we can assume that $x\geq 0$.
\begin{eqnarray}
 f(x)=\pi p_s(x,1,1)=\frac{1}{2}\int\limits_{\mathcal{R}}d k 
 \frac{e^{-i k x}}{1+|k|}\nonumber \\
=\int\limits_{0}^{\infty}d k \int\limits_{0}^{\infty}d s 
\cos{(k x)}e^{-s(1+k)}= \nonumber \\
= \int\limits_{0}^{\infty}d s \frac{s e^{-s}}{x^2+s^2}= 
\int\limits_{0}^{\infty}d t \frac{t e^{-t x}}{1+t^2}=\nonumber \\
-\frac{d}{d x} \int\limits_{0}^{\infty}d t \frac{e^{-t x}}{1+t^2}\equiv -
\frac{d}{d x} g(x).
\end{eqnarray}
One can verify that $g(x)$ is a solution of the equation
\begin{equation}
g''(x)+g(x)=\frac{1}{x},
\end{equation}
which is a second order inhomogeneous linear differential equation with 
constant coefficients. 
We can easily solve it by using the method of variation of parameters. 
Two constants in the general solution 
are calculated from the boundary conditions 
$g(0)=\frac{\pi}{2}$ and $\lim\limits_{x\to\infty}{g(x)}=0$. 
The solution reads:
\begin{equation}
g(x)=\left(\frac{\pi}{2}-\mathrm{Si}(x) \right)\cos{x}  + \mathrm{Ci}(x) \sin{x},
\end{equation}
which, after differentiation, leads to formula (\ref{Linnik1}).

 \section{Numerical scheme}
 \label{app-numerical-scheme}
The optimization problem at hand could not be solved analytically. We have thus 
solved it numerically. Scaling formulas Eq.(\ref{scaling}) allow for very fast 
numerical optimization, by reducing numerical calculation of the MFAT to one value 
of $x$ for each $\alpha$ and $r$. The algorithm then proceeds as follows: 
For each $\alpha$ we perform numerical integration by use of the reverse 
Fourier transform of the Linnik distribution, Eq.(\ref{Linnik}), for a 
given value $x$, e.g. $x=1$, and a few values of $r$. 
We fit a quadratic function to the calculated points and find the minimum 
of that function. Next we refine the interval of $r$ values, 
centering it at the estimated minimum and, consecutively, we reduce its 
length. This procedure is repeated until the desired accuracy is achieved. 
We end up with a quadratic function which, by means of its vertex coordinates, 
defines our $T^*_{\alpha} (1)$ and  $r^*_{\alpha} (1)$. 
Scaling equations, Eq.(\ref{scaling}), allow us to extend 
these results to arbitrary $x$.

When we start the calculation for a new value of $\alpha$, 
we face the problem of choosing a proper interval of values of $r$. 
Since we fit a quadratic function, 
it is important that the interval contains the optimal $r$. 
For this reason, we can make use of the optimal $r^*_{prev}$ that was 
calculated in a previous step, for a value of $\alpha$ close to the new one. 
Accordingly, we choose an interval of $r$ which contains $r^*_{prev}$. 
The formula for the optimal resetting frequency for the Brownian motion case 
is known \cite{majumdar2011resetting} and reads:
\begin{equation}
r^*_2(x)=\frac{D z^2}{x^2},
\label{opt_Brownian}
\end{equation}
with $z\approx1.5936$. Therefore, when performing numerical analysis, 
we have started our calculations from $\alpha=2$.
In the very last step we find, for each $x$, an $\alpha$-parameter for 
which the smallest optimal MFAT is obtained. This is our global minimum. 
The numerical scheme used for the optimization of 
$\langle T_{fa}(\lambda_t)\rangle$ is analogous.

\bibliography{citations}
\bibliographystyle{phjcp}
\end{document}